\documentclass[aps,final,notitlepage,oneside,onecolumn,nobibnotes,
nofootinbib,superscriptaddress,noshowpacs]{revtex4}

\usepackage{amssymb}
\usepackage{amsfonts}
\usepackage{bm}

\def\fun#1#2{\lower3.6pt\vbox{\baselineskip0pt\lineskip.9pt
\ialign{$\mathsurround=0pt#1\hfil ##\hfil$\crcr#2\crcr\sim\crcr}}}

\newcommand{\beq}{\begin{eqnarray}}
 \newcommand{\eeq}{\end{eqnarray}}
\newcommand{\be}{\begin{equation}}
 \newcommand{\ee}{\end{equation}}

\def\fun#1#2{\lower3.6pt\vbox{\baselineskip0pt\lineskip.9pt
\ialign{$\mathsurround=0pt#1\hfil ##\hfil$\crcr#2\crcr\sim\crcr}}}

\newcommand{{\SD}}{\rm SD}

\newcommand{\ver}{\mbox{\boldmath${\rm r}$}}

\newcommand{\vep}{\mbox{\boldmath${\rm p}$}}

\newcommand{\ves}{\mbox{\boldmath${\rm s}$}}

\begin{document}

\title{The Hyperfine Splittings in Heavy-Light Mesons and Quarkonia}

\author{\firstname{A.M.}~\surname{Badalian}}
\email{badalian@itep.ru}
\affiliation{Institute of Theoretical and Experimental Physics, Moscow, Russia}

\author{\firstname{B.L.G.}~\surname{Bakker}}
\email{blg.bakker@few.vu.nl} \affiliation{Department of Physics
and Astronomy, Vrije Universiteit, Amsterdam, The Netherlands}

\author{\firstname{I.V.}~\surname{Danilkin}}
\email{danilkin@itep.ru} \affiliation{Gesellschaft fur
Schwerionenforschung (GSI) Planck Str. 1, 64291 Darmstadt,
Germany}\affiliation{Institute of Theoretical and Experimental
Physics, Moscow, Russia}


\begin{abstract}
Hyperfine splittings (HFS) are calculated within the Field
Correlator Method, taking into account relativistic corrections.
The HFS in bottomonium and the $B_q$ (q=n,s) mesons are shown to
be in full agreement  with experiment if a universal coupling
$\alpha_{HF}=0.310$ is taken in perturbative spin-spin potential.
It gives $M(B^*)- M(B)=45.7(3)$~MeV, $M(B_s^*) -
M(B_s)=46.7(3)$~MeV ($n_f=4$), while in bottomonium
$\Delta_{HF}(b\bar b)=M(\Upsilon(9460))- M(\eta_b(1S))=63.4$~MeV
for $n_f=4$ and 71.1~MeV for $n_f=5$ are obtained; just latter
agrees with recent BaBar data. For unobserved excited states we
predict $M(\Upsilon(2S))-M(\eta_b(2S))=36(2)$~MeV,
$M(\Upsilon(3S))- M(\eta(3S))=28(2)$~MeV, and  also $M(B_c^*)=
6334(4)$~MeV, $M(B_c(2S))=6868(4)$~MeV, $M(B_c^*(2S))=6905(4)$
~MeV. The mass splittings between $D(2\,{}^3S_1) - D(2\,{}^1S_0)$,
$D_s(2\,{}^3S_1)-D_s(2\,{}^1S_0)$ are predicted to be $\sim 70$
MeV, which are significantly smaller than in several other
studies.
\end{abstract}

\maketitle

\section{Introduction}

Spin-spin interaction in mesons has been studied in a large number
of theoretical papers \cite{ref.1}-\cite{ref.6}, however, up to
now some characteristic features of this interaction are not fully
understood. This statement can be illustrated by theoretical
failure to explain two experimental facts: rather small
$\psi(3686)-\eta_c(2S)$ mass difference:
$M(\psi(3686))-M(\eta_c(2S))=49\pm 4$~MeV
\cite{ref.7}-\cite{ref.9} and, on the contrary, unexpected large
HFS in bottomonium, which follows from the mass
$M(\eta_b(1S))=9391.1\pm 3.1$~MeV of the $\eta_b(1S)$ meson,
recently discovered by the BaBar Collab. \cite{ref.10}. The
$\eta_b$ meson was observed in the radiative decays,
$\Upsilon(3S)\rightarrow \gamma\, \eta_b(1S)$  and
$\Upsilon(2S)\rightarrow \gamma\, \eta_b(1S)$ \cite{ref.10}, and
later confirmed by the CLEO Collaboration, also in the radiative
$\Upsilon(3S)\rightarrow\gamma\,\eta_b(1S)$ decay \cite{ref.11}.
The measured value, $\Delta_{HF}(b\bar b)=
M(\Upsilon(1S))-M(\eta_b(1S))=69.9\pm 3.1$~MeV, is significantly
larger than in most theoretical predictions, thus illustrating
that modern understanding of HF interaction in QCD remains
incomplete.

In perturbative approach a spin-spin potential between heavy
quarks contains the factors, like the strong coupling and quark
masses, which differ in different models and as a result,
theoretical predictions for $\Delta_{HF}(b\bar
b)=M(\Upsilon(9460))-M(\eta_b(1S))$ vary in wide range:
$35-90$~MeV \cite{ref.1}- \cite{ref.6}, \cite{ref.12}, being in
most cases smaller than experimental number.

On fundamental level spin-spin potential $V_{ss}$ has been
recently studied in quenched QCD on large lattice \cite{ref.13},
where this potential was shown to be compatible with zero at
distances $r\geq 0.30$ fm (for unknown reason at smaller $r$ it
has negative sign with a large magnitude). Although the lattice HF
potential remains undefined at small r, its behavior at larger $r$
is in agreement with widely used Fermi-Breit potential containing
$\delta^3(\vec{r})$ \cite{ref.14}. What is important that in
lattice QCD, as well as in Field Correlator Method (FCM)
\cite{ref.15}-\cite{ref.17}, a spin-spin potential is described by
universal functions, expressed via the field correlators.
Moreover, in \cite{ref.17} it was shown that nonperturbative HF
potential can give not small contributions to HFS.

On the other hand  in Ref. \cite{ref.1} a smearing procedure for
the $\delta^3(\vec{r})$-function was shown to be very important,
giving  a large Gaussian smearing parameter for heavy mesons,
containing a $b$ quark, so that for the $B_q~(q=n,s,c)$ mesons and
bottomonium a smearing occurs at very small distances and for them
the use of the Fermi-Breit potential may be a good approximation.
For the mesons, not containing a $b-$quark, a smearing parameter
is essentially smaller, both for light mesons and for the $D, D_s$
mesons \cite{ref.1}. However, such dependence of a smearing
parameter on a quark content does not agree with the lattice and
FCM representation about a spin-spin potential as a universal one
in static approximation, where is defined by the field correlators
with universal parameters \cite{ref.13}, \cite{ref.17}.

Recently HFS in the $B_q$ mesons and quarkonia have also been
calculated in lattice QCD \cite{ref.18}-\cite{ref.22} and their
results we shall shortly discuss in our paper. Here we study HFS
with the use of FCM, where both perturbative and nonperturbative
spin-spin potentials are presented in analytical form
\cite{ref.16}, \cite{ref.17} and it allows to analyse the role of
different physical parameters, defining HF structure. However, a
comparison of our and lattice results is rather difficult, because
in lattice calculations perturbative and nonperturbative spin-spin
effects are not separated and a characteristic value of the strong
coupling $\alpha_{HF}$ is not discussed. On the contrary, our
analysis shows that in bottomonium and the $B_q~(q=n,s,c)$ mesons
HFS can be described within only perturbative approach, since
nonperturbative spin-spin potential gives a small contribution.

It is important that in the $B_q (q=n,s)$ mesons and bottomonium a
good agreement with experiment is reached taking a universal
coupling, $\alpha_{HF}=0.310$ \cite{ref.23}. We would like to
stress that this number is significantly larger that that
prescribed in pQCD, where $\alpha_s(m_b)\sim 0.18$ and
$\alpha_s(m_c)\sim 0.23$ are used; just because of such small
coupling small HFS was predicted in bottomonium in \cite{ref.5},
\cite{ref.12}.

However, for charmonium and the $D, D_s$ mesons their HFS turn out
to be by $\sim 15\%$ and $\sim 30\%$ smaller than in experiment,
if the same $\alpha_{HF}=0.310$ is taken. It can occur for two
reasons: if for those states nonperturbative HF potential gives
essential contributions, or higher order corrections are not
small, as it takes place in fine structure splittings of the $1P$
charmonium multiplet \cite{ref.24}. The situation is different for
the charmonium excited states, for which just with a universal
coupling, $\alpha_{HF}\sim 0.31$, a good agreement with
experimental HFS: $M(\psi(3686))-M(\eta_c(3637))=49\pm 4$ MeV
\cite{ref.7} is obtained. Notice that the scale, corresponding to
$\alpha_{HF}(\mu)=0.31\simeq\alpha_s(\mu)$ is rather large,
$\mu\sim 1.7$ GeV.

In theoretical models two typical choices of $\alpha_{HF}$ are
used:

\begin{enumerate}
    \item
First one, when "a universal" $\alpha_{HF}$ is used. For example,
in \cite{ref.2}  $\alpha_{HF}=0.36$ was taken from the fit to the
mass difference, $M(J/\psi)-M(\eta_c(1S))=117$~MeV; then for this
choice predicted HFS in bottomonium, $M(\Upsilon(9460))-
M(\eta_b(1S)) = 87$~MeV, has appeared to be by $\sim 25\%$ larger
than experimental number. In \cite{ref.3} a smaller universal
$\alpha_{HF}=0.339$ was used in the heavy-light mesons; however,
it is difficult to compare our and their results, because in
\cite{ref.3} a large string tension, $\sigma=0.257$~$GeV^2$, was
used, while here (as well as in \cite{ref.1}) the conventional
$\sigma=0.18$~GeV$^2$ is taken. \item

Second choice is mostly used in pQCD \cite{ref.5}, \cite{ref.12},
where a scale $\mu=m_Q$ depends on a heavy quark mass and
therefore the value of $\alpha_{HF}(\mu)\simeq \alpha_s(m_Q)$ is
essentially smaller. In \cite{ref.12}, as well as in the EFG paper
\cite{ref.4}, just due to the choice of $\alpha_{HF}(m_b)=0.18$
small HFS were obtained in bottomonium (although the w.f. at the
origin from \cite{ref.12} have provided a precision description of
dielectron widths for $\Upsilon(nS)~(n=1,2,3)$ \cite{ref.25}).

\end{enumerate}

In FCM a spin-spin potential takes into account relativistic
corrections and the current masses are used for a light quark,
$m_n\sim 5$~MeV~$(n=u,d)$, and $m_s\simeq 200$ MeV for a $s$ quark
(about a choice of $m_s$ see \cite{ref.26}), so that the $B, D$,
and $B_s, D_s$ mesons can be considered on the same footing as
heavy quarkonia and the $B_c$ mesons.

We shall show here that HFS are sensitive to the value of the
vector QCD constant $\Lambda_V(n_f)$, defining a vector part of a
static potential in coordinate space. In  its turn this constant
is expressed via $\Lambda_{\overline{MS}}(n_f)$ \cite{ref.27},
which at present are known with a good accuracy only for $n_f=5$
and with $10\%$ accuracy for $n_f=3,4$ \cite{ref.7}. To fix
$\Lambda_V(n_f)$ we assume here, as well as in \cite{ref.1}, that
in the one-gluon-exchange potential (OGE) the freezing value of
the vector coupling $\alpha_V(r)(n_f)$ is the same for $n_f=3,4,5$
(it is denoted as $\alpha_{crit}$). Due to such an assumption the
HFS dependence on $n_f$ is weakening, with an exception of the
bottomonium ground states.

We also calculate HFS and the masses of  undiscovered yet mesons:
$\eta_b(2S)$, $\eta_b(3S)$, $B_c^*(1S)$, and the masses of
$B_q(2S), D(2S), D_s(2S)$.

The paper is organized as follows. In Section II the spin-spin
potential is given in the form, where relativistic corrections are
taken into account, as it is prescribed in FCM.  Also relativistic
string Hamiltonian is presented. In Section III the details of the
static potential are discussed. In Section IV calculated w.f. at
the origin and HFS for the $B_q$ mesons and bottomonium are given
and their dependence on the number of flavors is discussed. In
Section V a choice of the strong coupling for the $D, D_s$ mesons,
and charmonium is discussed. Conclusions of our analysis are given
in Section VI. In Appendix A the conventional formula for the pole
mass of a heavy quark is shortly discussed and in Appendix B we
describe the self-energy contribution to the meson mass, which is
important for heavy-light mesons.

\section{The HF potential in the Field Correlator Method}

The conventional form of the Fermi-Breit potential \cite{ref.13},

\be \hat V_{ss}(r) =\ves_1\ves_2 \frac{32\pi}{9} \frac{\alpha_{HF}
(\mu)}{\tilde m_1 \tilde m_2}\, \delta^{3} (\ver),\label{1}\ee
is widely used in heavy quarkonia, as well as in many
nonrelativistic models. It contains the constituent quark masses
$\tilde m_1$ and $\tilde m_2$, which are model-dependent and can
differ by $\sim 30\%$, or even larger, in different models, e.g.
the mass of a $c-$quark, $m_c=1.48$~GeV, was taken in
\cite{ref.2}, while in the Cornell potential  a larger  value,
$m_c=1.84$~GeV, was used \cite{ref.28}. A constituent mass is
supposed to be the same for all $nS$ and $nL$ states.

In Eq. (\ref{1}) the strong coupling $\alpha_{HF}(\mu)$ can differ
from the QCD strong coupling $\alpha_s(\mu)$ (in the
$\overline{MS}$ renormalization scheme) due to higher order
perturbative corrections. These higher order corrections in
one-loop approximation were calculated for heavy quarkonia
\cite{ref.29}:

\be \alpha_{HF}(\mu)=\alpha_s(\mu) \left[ 1+\frac{\alpha_s(mu)}
{\pi} \rho (n_f)\right], \label{2}\ee
but remain unknown for heavy-light mesons, containing a light (or
a strange) quark. Therefore in general case the coupling
$\alpha_{HF}$ in Eq.(\ref{1}) should be considered as an effective
one. Notice that its value is  smaller than a freezing constant of
the vector coupling $\alpha_V(r)$, which defines the OGE potential
at large distanses (or at small momenta) (see Eq.(\ref{17})). In
heavy quarkonia with $m_1=m_2=m_Q$ the factor $\rho$ is known
\cite{ref.29}:
\be
     \rho= \frac{5}{12}\beta_0 - \frac{8}{3} - \frac{3}{4}\ln2
\label{3} \ee
and appears to be small: $\sim 6-8\%$ for $n_f=3$, $\sim 3-4\%$
for $n_f=4$, and $\leq 0.1\%$ for $n_f=5$; still in some cases
these corrections can improve an accuracy of calculations.
However, since they are not defined for heavy-light mesons, here
in all cases we will consider $\alpha_{HF}$ as an effective
coupling, which is factually a fitting parameter.

The important role of relativistic corrections, even for the $B_c$
meson, has been underlined in \cite{ref.1}, \cite{ref.3}, and also
in the lattice calculations of $B_c^*$ in full QCD \cite{ref.21}.
In FCM relativistic corrections are taken into account in two
ways: firstly, through the kinetic energies of a quark and
antiquark, which enter a spin-spin potential \cite{ref.16},
\cite{ref.17}:
\be
    \hat V_{ss}(r) =\ves_1\ves_2 \frac{32\pi}{9}
    \frac{\alpha_{HF} (\mu)}{\omega_1\omega_2}\, \delta (\ver).\label{4}\ee
For this potential a HFS is
\be \Delta_{hf} (nS) =\frac{8}{9}
\frac{\alpha_{HF}(\mu)}{\omega_1\omega_2}|R_n(0)|^2,\label{5}\ee
where relativistic corrections are taken into account via the
averaged kinetic energies $\omega_1(nS), \omega_2(nS)$:
\begin{equation}
     \omega_1(nS)=\langle\sqrt{\mathbf{p}^2 + m_1^2}\,\rangle_{nS},~~
     \omega_2(nS) =\langle\sqrt{\mathbf{p}^2 +m_2^2}\,\rangle_{nS},\label{6}
\end{equation}
which are well defined. By definition they depend on the quantum
numbers of a given state $nS$, growing for larger $nS$ states. The
important point is that in (\ref{6}) the masses $m_1, m_2$ are not
arbitrary (or fitting parameters): they are equal the pole masses
$m_c$, $m_b$ in heavy quarkonia, which are now known with an
accuracy $\sim 70$~MeV for a $b$ quark and $\sim 100$~MeV for a
$c$ quark (see \cite{ref.7} and references therein). In leading
order the pole masses $m_Q$ do not depend on a number of flavors,
while in the order ($\alpha_s(\bar m_Q)^2$) they slightly depend
on $n_f$ (as in Eqs.(\ref{A.1}) and (\ref{A.2}) in Appendix A).
For heavy quarks we take the following pole masses:
$m_c=1.41$~GeV, $m_b=4.79$~GeV for $n_f=4$ and $m_b=4.82$~GeV for
$n_f=5$.

For a light quark $(n=u,d)$ we use the current mass  $m_n=5$~MeV
and $m_s=200$~MeV for a $s$-quark. The mass of a $s$ quark is
relatively large (close value of $m_s$ is used in the Dirac
equation in \cite{ref.3}), because the spectra of the $D_s, B_s$
mesons are defined at the scale, $\mu\leq 1$~GeV \cite{ref.26},
which is smaller than the conventional scale 2~GeV, for which
$m_s(2~GeV)\simeq 90$~MeV \cite{ref.7}.

For excited states the kinetic energies $\omega_i$ (\ref{6})
increase and therefore HFS, calculated with the HF potential
(\ref{4}), are always smaller than those for the Fermi-Breit
potential  (\ref{1}) with fixed (constituent) masses.

Other type of relativistic corrections enter via the w.f. at the
origin, which together with the $\omega_i$ are calculated from the
relativistic string Hamiltonian (RSH) $H_0$, also derived in FCM
\cite{ref.30},
  \begin{equation}
 H_0=\frac{\omega_1}{2} +\frac{\omega_2}{2} +\frac{m^2_1}{2\omega_1}+
 \frac{m^2_2}{2\omega_2} +\frac{\vep^2}{2\omega_{red}}
 +V_B(r).
\label{7}
\end{equation}
The variables  $\omega_i$ enter $H_0$ as the kinetic energy
operators. However, while a HF interaction (as well as any
spin-dependent potential) is considered  as a perturbation, then
in (\ref{4}), (\ref{5}) $\omega_i$ should be changed by the matrix
elements (m.e.) (\ref{6}) \cite{ref.16}.

>From RSH a simple expression follows for a spin-averaged mass
$M(nS)$   \cite{ref.31}:
\begin{equation}
 M(ns)=\frac{\omega_{1}}{2}+\frac{m_1^2}{2\omega_1}+\frac{\omega_{2}}{2}+\frac{m_2^2}{2\omega_1}
 + E_{nS}(\omega_{red}).
\label{8}
\end{equation}
In (\ref{8}) an excitation energy $E_{nS}(\omega_{red})$ depends
on the reduced mass:
$\omega_{red}=\frac{\omega_1\omega_2}{\omega_1 +\omega_2}$. Also
in bottomonium the mass formula (\ref{8}) does not contain any
overall constant, while for heavy-light mesons  a negative (with
not small magnitude) self-energy term, proportional to
$(\omega_q)^{-1}$ ($q=n,s$), should be added \cite{ref.32} (see
the expression (\ref{B.1}) in Appendix B).

We use here the Einbein Approximation (EA), when  the variables
$\omega_i(nS)$, the excitation energy $E_{nS}(\omega_{red})$, and
the w.f.  are calculated from the Hamiltonian (\ref{6}) and two
extremum conditions, $\frac{\partial M(nS)}{\partial\omega_i}=0$
$(i=1,2)$, which are put on the mass $M(nS)$ \cite{ref.31}:
\begin{equation}
 \left[\frac{\omega_1}{2}+\frac{\omega_2}{2} +\frac{m_1^2}{2\omega_1} + \frac{m^2_2}{2\omega_2}+
 \frac{\bm{p}^2}{2\omega_{red}}+V_B(r) \right]\varphi_{nS}(r)
  = E(nS)\,\varphi_{nL},
\label{9}
\end{equation}

\be \omega_i^2(nS)  = m_i^2 -2\omega_i^2\,\frac{\partial E(nS,
\mu_{red})}{\partial \omega_i(nS)}\quad (i=1,2).\label{10}\ee
Before to define the w.f. at the origin and $\omega_i(nS)$  in
next Section we shortly discuss the static potential $V_B(r)$ in
the Eq.(\ref{9}).

\section{The Static Potential $V_B(r)$}

In a Hamiltonian approach a choice of a static potential $V_B(r)$
is of a special importance; we take it as a sum of linear
confining term and the OGE -type term: this additivity of a static
potential is well established now in analytical studies
\cite{ref.33} and on lattice \cite{ref.34}, \cite{ref.35}:
 \be
   V_B(r)=\sigma\ r +\frac{4\alpha_B(r)}{3\ r}.\label{11}\ee
For the string tension we use the conventional value,
$\sigma=0.18$ GeV$^2$, for all mesons (if their sizes are less
than $\sim 1$ fm. This our choice is in contrast to that in
\cite{ref.3}, where in the Dirac equation large
$\sigma=0.257$~GeV$^2$ was used for heavy-light mesons.

The OGE term  contains the vector coupling $\alpha_V(r)$, taken
here in a particular case from the background perturbation theory
(BPT) and denoted as $\alpha_B(r)$ \cite{ref.36}, \cite{ref.37}.
Two important conditions have to be put on a vector coupling:
\begin{enumerate}
\item
  i) As in pQCD, it has to possess the asymptotic freedom (AF)
property; just due to this property a static interaction depends
on a number of flavors. Also the AF behavior strongly affects the
w.f. at the origin.\item

 ii) The vector coupling freezes at large distances. The property
of freezing was widely used in phenomenology
\cite{ref.1}-\cite{ref.3} and confirmed in lattice calculations of
a static potential \cite{ref.34}, where a freezing property was
assumed  at rather small quark-antiquark separations, $r\geq 0.2$
fm. \end{enumerate}

On phenomenological level the freezing phenomenon has been
suggested in \cite {ref.38}, where in momentum space the logarithm
$\ln\frac{q^2}{\Lambda^2}$ in $\alpha_s(q^2)$ was changed by
$\ln\frac{q^2+4m_g^2}{\Lambda^2}$, thus introducing a regulator
$4m_g^2$. The mass $m_g$ was interpreted as an effective gluon
mass, although a meaning of $m_g$ is not well defined, since in
QCD a gluon has no a mass. Later a freezing phenomenon was studied
in BPT \cite{ref.37}, where this regulator was shown to be equal a
mass of the lowest hybrid excitation (called the background mass),
with $M_B=1.0\pm 0.05$~GeV \cite{ref.39} for $n_f=4,5$ and a
larger value, $M_B\sim 1.5$~GeV for $n_f=0$ \cite{ref.36}. As in
\cite{ref.1}, we shall call a freezing constant a critical one and
denote it as $\alpha_{crit}$.

Unfortunately, the critical constants, calculated from the static
potentials on lattice, are significantly smaller than those in
phenomenology and BPT. Here in BPT we use rather large
$\alpha_B(crit)=0.58-0.60$ for $n_f=4,5$, which are close to
$\alpha_{crit}=0.60$ (for any $n_f$) taken in  \cite{ref.1}. On
lattice the freezing effect occurs at small distances, $r\geq 0.2$
fm, and small $\alpha_{crit}(lat)\simeq 0.30$  in full QCD
($n_f=3$) \cite{ref.35} and $\alpha_{crit}(lat)\simeq 0.22$ in
quenched calculations ($n_f=0$) \cite{ref.34} were obtained. The
reasons of these discrepancies are not established yet.

While the critical value $\alpha_{crit}$ is fixed, then  with the
use of Eq.(\ref{17}) the constant $\Lambda_B(n_f)$ (for a given
$n_f$) can be defined. It is important that this constant cannot
be considered as a fitting parameter, because it is expressed via
the QCD constant $\Lambda_{\overline{MS}}(n_f)$ in the
${\overline{MS}}$ renormalization scheme \cite{ref.27} (see the
relation (\ref{16})). Therefore one can state that
$\Lambda_{\overline{MS}}$ indirectly defines $\alpha_{crit}$.

In the OGE term (\ref{11}) a vector coupling in coordinate space
$\alpha_B(r)$ is defined through the vector coupling
$\alpha_B(q^2)$ in the momentum space \cite{ref.25},
\cite{ref.36}:
\begin{equation}
\alpha_B(r) =\frac{2}{\pi}\int\limits_0^\infty
dq\frac{\sin(qr)}{q}\,\alpha_B(q). \label{12}
\end{equation}
Here the vector coupling  $\alpha_B(q^2)$ is taken  in two-loop
approximation,
\begin{equation} \alpha_B(q) =\frac{4\pi}{\beta_0t_B}\left(1-\frac{\beta_1}{\beta_0^2}
  \frac{\ln t_B}{t_B}\right),
\label{13}
\end{equation}
where the logarithm,
\begin{equation}
 t_B=\frac{q^2+M_B^2}{\Lambda_B^2},
\label{14}
\end{equation}
contains the constant  $\Lambda_B(n_f)$ defined via the QCD
constant $\Lambda_{\overline{MS}}(n_f)$. The relation between them
has been established in \cite{ref.27}:
\begin{equation}
   \Lambda_B(n_f)=\Lambda_{\overline{MS}}\exp\left(-\frac{a_1}{2\beta_0}\right),
   \label{15}
   \end{equation}
with $\beta_0=11 -\frac{2}{3}n_f$ and
$a_1=\frac{31}{3}-\frac{10}{9}n_f$. From the relation (\ref{15})
one can see that for a given $n_f$ the constant $\Lambda_B(n_f)$
is significantly larger than $\Lambda_{\overline{MS}}$:
\begin{eqnarray}
\nonumber \Lambda_B^{(5)}=1.3656\Lambda_{\overline{MS}}^{(5)}\quad (n_f=5);\\
\nonumber \Lambda_B^{(4)}=1.4238\Lambda_{\overline{MS}}^{(4)}\quad (n_f=4);\\
\Lambda_B^{(3)}=1.4753\Lambda_{\overline{MS}}^{(3)}\quad (n_f=3).
\label{16}
\end{eqnarray}

At present the QCD constant $\Lambda_{\overline{MS}}^{(5)}$  (for
$n_f=5$) is known from experimental value of
$\alpha_s(M_Z)=0.1182\pm 0.0012$ \cite{ref.7}. Then in two-loop
approximation it gives
$\Lambda_{\overline{MS}}(\textrm{two-loop})=232(12)$~MeV. The QCD
constants  $\Lambda_{\overline{MS}}$ for $n=3,4$ are extracted
from experiments with lower accuracy, $\sim 10\%$ \cite{ref.7}. To
define them we fix here the freezing value $\alpha_{crit}$,
assuming that they are the same for $n_f=3,4,5$. Then from
$\alpha_{crit}$ (\ref{17}) one can calculate all constants
$\Lambda_B(n_f)$ and then from (\ref{16}) to define
$\Lambda_{\overline{MS}}(n_f)$.

Notice that the mass $M_B$ depends on $\sigma$, being proportional
to $\sqrt{\sigma}$, and for $\sigma=0.18$ GeV$^2$ the value,
$M_B=1.0\pm 0.05$~GeV, was extracted from a detailed comparison of
the static force in FCM and lattice QCD \cite{ref.39} and also the
analysis of the bottomonium spectra \cite{ref.40}. Here we use
$M_B=0.95$~GeV. We also use here two values of
$\Lambda_{\overline{MS}}^{(5)}$, equal  236~MeV and 245~MeV, which
give the critical couplings 0.58 and 0.605.

From (\ref{12}) it can be easily shown that the critical couplings
in momentum and coordinate space coincide,
$\alpha_B(crit)=\alpha_B(q^2=0)= \alpha_B(r\rightarrow \infty)$,
and it is given by the expression \cite{ref.36}:
\begin{equation}
   \alpha_B(crit)=\alpha_B (r\to \infty) =\alpha_B(q=0) =\frac{4\pi}{\beta_0
   t_0} \left( 1-\frac{\beta_1}{\beta_0^2}\frac{\ln
   t_0}{t_0}\right), \label{17}
\end{equation}
with  $t_0= t_B(q^2=0)=ln\left(\frac{M_B^2}{\Lambda_B^2}\right)$.

From (\ref{16}) and  $\Lambda_{\overline{MS}}^{(5)}=245 (236)$~MeV
one obtains that in two-loop approximation $\alpha_s(M_Z)=0.1194
(0.1188)$, which agrees within an error with the world average
$\alpha_s(M_Z)=0.1184\pm 0.0012$ \cite{ref.7}. In Table I we
summarize the values of $\Lambda_B, \Lambda_{\overline{MS}}$ for
$n_f=3,4,5$.

\begin{table}
\caption{The vector constants $\Lambda_B$ and
$\Lambda_{\overline{MS}}$ (in MeV) ($n_f=3,4,5$) for
$\alpha_{crit}=0.605$.\label{tab.1}}

\begin{tabular}{llll}
\hline\hline
    $n_f$           &        3      &    4   &       5\\
\hline
    $\Lambda_B$          &     400    &    372     &   335\\
$\Lambda_{\overline{MS}}$  &   271  &      261    &    245\\
\hline\hline
\end{tabular}
\end{table}

The solutions of  the coupled equations (\ref{9}), (\ref{10}),
like the spectra and w.f at the origin, have been checked in
numerous studies \cite{ref.23}-\cite{ref.26}, \cite
{ref.39}-\cite{ref.40}, demonstrating a good description of
different physical characteristics.

\section{Relativistic Corrections and the wave functions at the origin}

With the use of the Eqs. (\ref{9}) and (\ref{10}) the kinetic
energies $\omega_i$ are calculated (see Table II).  Their numbers
show that relativistic corrections are small  in bottomonium:
$\omega_b(1S)- m_b \simeq 180$~MeV ($\sim 4\%$) and a bit larger,
$\sim 7\%$, for the $2S, 3S$ states. It is of interest to notice
that a relativistic correction to the $b-$quark mass in the $B_c$
meson is even smaller than in bottomonium, $\omega_b(1S)
-m_b\simeq 110$~MeV ($\sim 2\%$), while for a $c-$quark in $B_c$
the difference, $\omega_c(1S)- m_c\simeq 340$~MeV, is already
$\sim 25\%$ (all numbers refer to $\alpha_{crit}=0.605$,~$n_f=4$).

For the $B_q$ mesons  ($q=n,s,c$) relativistic corrections,
$\omega_q(1S)-m_q$  are given in Table II.\\

\begin{table}
\caption{The kinetic energies $\omega_q(1S) (q=n,s,c)$ and
$\omega_b(1S)$ (in MeV) for the static potential $V_B(r)$
(\ref{11}) with $n_f=4$,  $\Lambda_B=372$ MeV,
$\alpha_{crit}=0.605$. \label{tab.2}}
\begin{tabular}{llll}
\hline\hline
Meson            &   $B$  &   $B_s$ &  $B_c$\\
\hline
$ m_q $          &    5   &   200   &  1410 \\
$\omega_q  -m_q$ &   633  &   486   &  340\\
$m_b$       &  4790  &  4790   &  4820\\
$\omega_b -m_b $ &    42  &   46    &  110 \\
\hline\hline
\end{tabular}\\
\end{table}

In charmonium relativistic corrections to a $c-$quark mass are
already $\sim 13\%$ for the ground state and $\sim 17\%$ for the
$2S$ state. It is of interest to notice that they are even
smaller, $\sim 7\%$ for a $c$ quark in the $D, D_s$ mesons (see Table III).\\

\begin{table}
\caption{The kinetic energies $\omega_q(nS) (q=n,s)$ and
$\omega_c$ (in~MeV) for the static potential $V_B(r)$ $(n_f=4)$
with the same parameters as in Table II.\label{tab.3}}
\begin{tabular}{lllll}
\hline\hline
   Meson    &       $D $        &$ D_s$  &   $c\bar c(1S)$ &  $c\bar c(2S)$\\
\hline
   $ m_q$   &        5  &        200 &            1410   &      1410\\

$\omega_q -m_q$       &   542    &     400&         184           &   245\\

$\omega_c- m_c$    &     102   &   108  &            184&            245\\
\hline\hline
\end{tabular}
\end{table}

For the analysis of HFS it is convenient to introduce the ratio
$g_{B_q}, g_{D_q}$,  and also $g_b\equiv g(b\bar b), g_c\equiv
g(c\bar c)$,
 \be
 g_{B_q}(nS) = \frac{|R_n(0)|^2}{\omega_1(nS)\,\omega_2(nS)},\label{18}\ee
which directly enters the HFS (\ref{4}):
 \be
 \Delta_{HF}(nS)=\frac{8}{9}\,\alpha_{HF}(\mu)\, g_{B_q}(nS)\label{19}\ee
and appears to be weakly dependent on small variations of the
masses $m_1, m_2$, and other parameters of a static potential,
which are compatible with a good description of the meson
spectrum. Their values for the $B_q$ mesons and bottomonium are
given in Tables IV and V.

In general case the w.f. at the origin is sensitive to the choice
of $\Lambda_B(n_f)$, defining the OGE term, and may be $\sim 1.5$
times larger, if the AF behavior is neglected \cite{ref.28}. In
our calculations, when the AF property is taken into account and
the same $\alpha_{crit}(n_f)$ is used for $n_f=3,4,5$, the
differences in the $g_{B_q}, g_{D_q}$ values turn out to be small,
$\leq 5\%$ for  different $n_f$, (see Table IV), with exception of
$g_b$ in bottomonium.

Below we give the values of $g_b(n_f)$ for $n_f=3,4,5$, which were
calculated with the same $\alpha_{crit}=0.605$ . For this
$\alpha_{crit}$ and $M_B=0.95$ GeV in two-loop approximation the
constants $\Lambda_B(n_f=3)=0.40$ GeV,
$\Lambda_B(n_f=4)=0.372$~GeV, and $\Lambda_B(n_f=5)=0.335$ GeV are
easily calculated (see Table I). The largest difference takes
place in bottomonium, where
\be  g_b(n_f=3)=0.213~ \textrm{GeV},\quad   g_b(n_f=4)=0.230
~\textrm{GeV},\quad
       g_b(n_f=5)=0.258 ~\textrm{GeV},\label{20}\ee
i.e. $g_b(n_f)$ can change by $\sim 20\%$.  Therefore, if
experimental HFS in bottomonium are known with a precision
accuracy, then one can distinguish between cases with different
$n_f$. From the numbers given in (\ref{20}) and taking
$\alpha_{HF}=0.310$, as for the $B_q$ mesons, one obtains the
following HFS $\Delta_{HF}(b\bar
b)=M(\Upsilon(9460))-M(\eta_b(1S))$:
\be
  \Delta_{HF}(n_f=3)=58.7\, \textrm{MeV};\quad   \Delta_{HF}(n_f=4)=63.4\,
  \textrm{MeV};\quad
    \Delta_{HF}(n_f=5)=71.1\, \textrm{MeV} \label{21}\ee
Notice that our value for the $\Upsilon(9460)-\eta_b(1S)$ mass
difference with $n_f=3$ has appeared to be in good agreement with
the lattice calculations, also performed with  $n_f=3$, and where
$\Delta_{HF}(b\bar b)=61\pm 13\pm 4$~MeV was obtained in
Ref.\cite{ref.18}; $70\pm 11$ MeV in \cite{ref.19}, and a smaller
splitting, $\Delta_{HF}(b\bar b)= 54\pm 12$~MeV, was calculated in
\cite{ref.20}. (An accuracy of our calculations for HFS is
estimated to be $\pm 4$~MeV ($\sim 5\%$), as it follows by varying
different parameters, i.e. is better than in lattice calculations,
where at present it is $\geq 20\%$).

For the $B_q$ mesons their w.f. at the origin, the factors
$g_{B_q}$, and the HFS are given in Table IV for $n_f=3,4$.
\begin{table}
\caption{The ratios $g_{B_q}$ (\ref{18})(in~GeV) and $|R_1(0)|^2$
(in GeV$^3$) for the $B_q(1S)$ mesons ($\alpha_{crit}=0.605,
\Lambda_B(n_f=4)=372$~MeV,
$\Lambda_B(n_f=3)=400$~MeV).\label{tab.4}}
\begin{tabular}{llll}
\hline\hline
                    &     $ B  $ &        $ B_s $  &      $ B_c$\\
\hline
$|R(0)|^2(n_f=4) $&      0.510   &     0.558   &   1.742 \\

$ g_{B_q}(n_f=4) $&       0.165 &     0.168  &     0.212\\

$|R(0)|^2(n_f=3) $  &     0.478 &      0.527   &    1.626\\

$ g_{B_q}(n_f=3) $&       0. 158 &      0.161   &    0.198\\
\hline\hline
\end{tabular}
\end{table}
As seen from Table IV, for the $B_q$ mesons a difference between
$g_{B_q}(n_f=3)$ and $g_{B_q}(n_f=4)$ appear to be only by $\sim
4\%$; also for a given $n_f$ a difference between $g_{B}$ and
$g_{B_s}$ is small, $\leq 2\%$.

In  bottomonium a difference between $g_b(n_f=3)$ and $g_b(n_f=4)$
is larger, $\sim 12\%$, and in both cases they are smaller than
$g_b(n_f=5)$; corresponding HFS are given in Table V.
\begin{table}
\caption{The ratios $g_b(nS)$ (in~GeV), $|R_n(0)|^2$ (in GeV$^3$),
and HFS $\Delta_{HF}(nS)$ (in~MeV)  for the $1S, 2S, 3S$
bottomonium states ($n_f=4,5$,
$\alpha_{crit}=0.605$).\label{tab.5}}
\begin{tabular}{llll}
\hline\hline
                           & 1S &        2S &        3S\\
\hline
$|R_n(0)|^2 (n_f=5)$  &     6.476 &       3.398  &   2.682\\

  $g_b  (n_f=5) $ &         0.258 &        0.134   &   0.105\\

$\Delta_{HF}(n_f=5)$&         71.1 &         36.9&    28.9 \\

 $|R_n(0)|^2  (n_f=4)$&    5.668  &        3.126 &    2.508\\

  $  g_b   (n_f=4) $&      0.230&          0.127  &    0.100\\

$\Delta_{HF}(n_f=4)$&        63.4 &          35.0 &     27.6\\
\hline\hline
\end{tabular}
\end{table}
For excited states  the $\Upsilon(nS)-\eta_b(nS)$ splittings
($n=2,3$) are given also for the coupling $\alpha_{HF}=0.310$ and
this choice seems to be a realistic, because in bottomonium a
characteristic momentum weakly changes for excited states:
\be
\omega_b(1S)- m_b =185(5) \textrm{MeV},\quad  \omega(2S)-m_b=
195(5) \textrm{MeV},\quad   \omega_b(3S)-m_b=225(5)
\textrm{MeV}.\label{22}
\ee
In (\ref{22}) theoretical errors given come from slightly
different pole mass of a $b$ quark for different $n_f$.

Finally in Tables VI, VII we give $g_{D}, g_{D_s}$ for ground
states, and also $g_c(nS)\equiv g_{c\bar c}(nS)$ for the $1S$ and
$2S$ charmonium states.
\begin{table}
\caption{The w.f. $|R(0)|^2$ (in GeV$^3$) and the ratios
$g_{D},~g_{D_s},~g_c(1S),~g_c(2S)$ (in~GeV) for the $D, D_s$
mesons and charmonium (in the static potential $V_B(r)$
$\sigma=0.18$~GeV$^2$, $\Lambda_B(n_f=3)=0.40$~GeV,
$\Lambda_B(n_f=4)=0.372$~GeV, $M_B=0.95$~GeV).\label{tab.6}}
\begin{tabular}{lllll}
\hline\hline
              &  $D(1S)$ &  $D_s(1S)$ &   $c\bar c(1S)$&   $c\bar c(2S)\footnote{For the $2S$ charmonium state in the w.f. at the origin
the $S-D$ mixing between $\psi(3686)$ and $\psi(3770)$ states  is taken into account}$\\
\hline
$|R(O)|^2$    & 0.314   &  0.346 &    0.856 &   0.476\\

g(nS)  &     0.379  &    0.379  &     0.340   & 0.174\\
\hline\hline
\end{tabular}
\end{table}
It is worthwhile to notice that the factor $g_{c\bar c}(2S)$ is
two times smaller than $g_{c\bar c}(1S)$, being one of the reasons
why the mass difference  $\psi(3686)-\eta_c(2S)$ is small.

We would like to remind here that while the relativistic string
Hamiltonian (\ref{7}) is used, for a meson excited states may be
considered on the same footings as a ground state till, if a
single-channel approximation can be applied. Otherwise one has to
use the multichannel Hamiltonian, also derived within FCM
\cite{ref.41}.

\section{HFS in bottomonium and the $B_q$ mesons}

Firstly, for bottomonium we compare calculated here HFS, Eq.
(\ref{21}), with experimental HFS, which follows from experimental
mass of $\eta_b(1S)$:
\begin{eqnarray}
\nonumber &&\Delta_{HF}(b\bar b) =M(\Upsilon(9460))- M(\eta_b)=69.9\pm 3.2\,
\textrm{MeV}\quad [10];\\ && \Delta_{HF}(b\bar b)= 68.5\pm 6.6\,
\textrm{MeV}\quad [11]. \label{23}
\end{eqnarray}
In the BaBar experiment the $\eta_b$ mass is defined with a small
errors, $\pm 3.2$~MeV, and even with a smaller error, $\leq
1$~MeV, the mass differences: $M(B^*)-M(B)$, $M(B_s^*)-M(B_s)$ are
known now from experiment \cite{ref.7}. Such good experimental
data allow to extract a coupling $\alpha_{HF})$ with a good
accuracy. For fitting procedure the important point is that in
bottomonium a fitted value of $\alpha_{HF}$ cannot be larger that
that for $B_q$ mesons, since due to the AF property a coupling of
a smaller size system (bottomonium) is typically smaller. For the
$B_q$ mesons and bottomonium the best fit is obtained for
$\alpha_{HF}=0.310$, later called "a universal" coupling.

As seen from (\ref{21}), in bottomonium full agreement with
experiment is reached if  $\alpha_{HF}=0.310$ and $n_f=5$ are
taken in the static potential. It gives $\Delta_{HF}(b\bar
b)=71.1$~MeV, which is by 8 MeV larger than $\Delta_{HF}(b\bar
b)=63.4$~MeV for $n_f=4$. (We do not consider as unphysical a fit
with $n_f=4$ and $\alpha_{HF}=0.348$ when agreement with
experiment is also possible, since the value of this coupling is
by $12\%$ larger than that for the $B_q$ mesons). Notice  that the
best description of the bottomonium spectra takes place also for
$n_f=5$ \cite{ref.40}.

In Table V the HFS for the bottomonium $2S, 3S$ states are given;
they are equal 36(1)~MeV and 28(1)~MeV, respectively, being weakly
dependent on $n_f$ taken.

The HFS of the $B_q$ ground states are presented in Table VII,
while in Table VIII  the mass splttings for higher states,
$B(2S),~B_s(2S),~B_c(2S)$, are also given.

\begin{table}
\caption{The HFS (in~MeV) in the $B_q$ mesons with
$\alpha_{HF}(n_f=4)=0.310$, $\Lambda_B(n_f=4)=0.372$~GeV and
$\alpha_{HF}(n_f=3)=0.324$, $\Lambda_B(n_f=3)=0.40$
GeV.\label{tab.7}}
\begin{tabular}{llll}
\hline\hline
   HFS &            $ B  $ &      $ B_s $&    $ B_c(1S)$\\
\hline

$\Delta_{HF}(n_f=4)$&   45.6             &  46.3      &   58.4 \\

$\Delta_{HF}(n_f=3)$ &   45.4            46.1  &    57.2\\

$\Delta_{HF}(\exp) $&   $45 .78\pm 0.35$   &  $46.5\pm 1.25$  &   abs \\
\hline\hline
\end{tabular}
\end{table}

\vspace{1cm}

For the $B^*(1S)-B(1S), B_s^*(1S)-B_s(1S)$ splittings a good
agreement with experiment is reached in two cases: with $n_f=4$
and $\alpha_{HF}=0.310$ (as in bottomonium) and with $n_f=3$ and a
bit larger $\alpha_{HF}=0.324$ (see Table VII). First choice seems
to be preferable as a universal one, but in any case a difference
between two couplings is small, $\leq 5\%$. Therefore one can
speak about a universal $\alpha_{HF}$ within $5\%$ accuracy. To
fix a preferable number $n_f$ for the $B, B_s$ mesons one needs to
use an additional information, like the decay constants etc.

For the $B_c$ mesons our splitting, $M(B_c^*)-M(B_c)=58.4$~MeV for
$n_f=4$ and 57.2 MeV for $n_f=3$, appears to be in agreement with
the unquenched lattice calculations from \cite{ref.21}, where the
number $53\pm 7$~MeV is predicted.

For excited $B_c(2S)$ states our calculations give the centroid
mass, $M_{cog}(B_c(2S))=6893$~MeV, and the $B_c^*(2S)-B_c(2S)$
splitting 37.3~MeV (see Table VIII), from which
$M(B_c^*(2\,{}^3S_1))=6.902$~MeV and
$M(B_c(2\,{}^1S_0))=6865$~MeV. An accuracy of our calculations,
performed in single-channel approximation, is estimated to be $\pm
5$ MeV , although for
higher states an influence of open channel(s) may be important.\\

\begin{table}
\caption{The masses $M(B_q(2S))$, $M(B_q^*(2S))$, and the HFS
$M(B_q^*(2S)-M(B_q(2S)),~(q=n,s,c)$ (in~MeV) for
$\alpha_{HF}(n_f=4)=0.310$ and $\alpha_{crit}(n_f=4)=0.605$.
\label{tab.VIII}}
\begin{tabular}{llll}
\hline\hline
 Meson       &             $ B(2S) $&   $B_s(2S)$&         $ B_c(2S)$\\
\hline
  $g_{B_q}(2S)$&          0.124&    0.1255&        0.1353 \\

 $\Delta_{hf}(B_q(2S))$&  34.3  &  34.7&       37.3 \\

    $M(B_q(2\,{}^1S_0))$&          5967&     6040&        6868\\

 $M(B_q^*(2\,{}^3S_1))$&           6001&     6075&         6905\\
 \hline\hline
\end{tabular}
\end{table}

From Table VIII one can see that for excited $B, B_s, B_c$ mesons
their HFS have close values, $\Delta_{HF}(B_q(2S)\sim 34-37$~MeV.
and the masses of singlet and triplet $2S$ states are also given
in Table VIII. We would like to notice that our HFS for the $B_s$
mesons differ from those calculated in lattice QCD \cite{ref.22},
where  small $\Delta_{HF}(B_s(1S))=29.8\pm 3.2$~MeV was calculated
for the ground $1S$ states (in our calculations it is equal
46.3~MeV), while on the contrary, in \cite{ref.22} for the excited
$2S$ states a central value of the HFS,
$\Delta_{HF}(B_s(2S))=56\pm 27$~MeV, is larger than in our
calculations, where this splitting is only 35~MeV.

Our calculations have been performed in single-channel
approximation with a string tension $\sigma=const=0.18$ GeV$^2$.
For higher levels a influence of open channels can be taken into
account, e.g. via a flattening of the static potential, and then
the masses $M(B(2S)), M(B_s(2S))$ appear to be only by $\sim
10$~MeV smaller. The effect from open channels may be more
important for the $D(2S)$, $D_s(2S)$ states and then a
multichannel relativistic Hamiltonian from \cite{ref.41} may be
used instead of the equation (\ref{7}).

Thus our analysis of HFS for the $B(1S), B_s(1S)$ mesons and
bottomonium shows that a good agreement with experiment is reached
with a universal $\alpha_{HF}=0.310$. This coupling is smaller
than that in \cite{ref.2}, \cite{ref.3} and corresponds to rather
large renormalization scale, $\mu\simeq 1.70$ GeV. This value of
the scale confirms existing interpretation of the spin-spin
potential as dominantly a short-range perturbative one, thus
justifying the use of the $\delta(\ver)$-function.

\section{Large HFS in charmonium and the $D,~D_s$ mesons}

Experimental HFS for the $D,~D_s$ ground states are large, $\simeq
140$~MeV, being three times larger than those for the $B,~B_s$
mesons. Let us firstly estimate these HFS, taking the factors
$g_D=g_{D_s}=0.379$~GeV from Table VI ($n_f=4$), and the value
$\alpha_{HF}=0.310$, as for the $B_q$ mesons. Then for the ground
states we obtain
$\Delta_{HF}(D(1S))=\Delta_{HF}(D_s(1S))=104.4$~MeV, which are by
$35\%$ smaller than experimental HFS \cite{ref.7}.

In charmonium a discrepancy between calculated HFS with
$\alpha_{HF}=0.310$, $\Delta_{HF}(1S,c\bar c)= 93.7$ MeV,and
experimental number is smaller, $\sim 20\%$. If in charmonium
first order correction (\ref{3}), equal $\leq 6\%$, is taken into
account, then this discrepancy remains not small, $\sim 15\%$.

For the $2S$ charmonium states, a situation is different and full
agreement with experimental HFS takes place, if a universal
coupling $\alpha_{HF}=0.31$ is used. With the factor
$g_c(2S)=0.174$~GeV from Table VI, one obtains
\be
 \Delta_{HF}(2S,c\bar c)=M(3686)-M(\eta_c(2S))=47.9, \, \textrm{MeV}\ee
coinciding with the experimental $\psi(3686)-\eta_c(2S)$ mass
difference: $\Delta_{HF}(2S,c\bar c)|_{exp}= 48\pm 4$ MeV
\cite{ref.7}. (To get this result we have taken into account the
$S-D$ mixing between $\psi(3686)$ and $\psi(3770)$ with the mixing
angle $\theta\sim 11^\circ$ \cite{ref.42}). Thus for the $2S$
charmonium states a universal coupling provides agreement with
experiment. Therefore we expect that for the $2S$ states of other
heavy-light mesons the coupling $\alpha_{HF}=0.31$ can be also
used.

However, it remains unclear what kind of corrections ($\sim 15\%$
in charmonium ground states) have been lost in our analysis? (We
remind that first order perturbative contribution gives only $\leq
6\%$.) We assume here that such a contribution comes from
nonperturbative spin-spin potential, and just due to
nonperturbative correlators lattice calculations
\cite{ref.18}-\cite{ref.20} give the mass difference,
$M(J/\psi)-M(\eta_c(1S))$, in good agreement with experimental
number, equal 117~MeV. Such HFS, coming from nonperturbative
spin-spin potential, can be taken into account also within FCM,
where recently new results have been obtained for the vacuum
correlation functions and correlation lengths \cite{ref.17},
\cite{ref.43}. Detailed analysis of these effects will be
considered in our next paper.

However, we can use results from Ref.\cite{ref.25}, thatfor higher
states  nonperturbtive contributions are much smaller than for the
ground states; just for that reason we have obtained a good
description of HFS for the charmonium $2S$ states with
$\alpha_{HF}=0.31$. Therefore we use here only perturbative part
of HFS for higher states.

It is of interest to notice that for $D(2S)$ and $D_s(2S)$ their
w.f. at the origin and the quark kinetic energies can differ by
$\sim 10\%$ , nevertheless the factors $g_D(2S)$ and $g_{D_s}(2S)$
coincide within $4\%$ accuracy, being equal $0.273\pm 0.005$~GeV,
if linear confining term is taken in the static potential; it
gives $\sim 75$ MeV for their HFS . If one takes into account
flattening of confining potential, which is often important for
higher states \cite{ref.44}, then a small decreasing of this
factor takes place: $g_D(2S)\simeq g_{D_s}(2S)=0.264\pm 0.004$
GeV, where the error occurs due to possible different choice of
$n_f=3$ or $n_f=4$; in this case the HFS are equal $72\pm 3$~MeV
for the $D(2S),~D_s(2S)$ states. Calculated HFS are presented in
Table IX.\\

\begin{table}
\caption{The HFS (in~MeV) of the $D(2S)$, $D_s(2S)$ mesons, and
charmonium with $\alpha_{HF}=0.31$; experimental HFS from
\cite{ref.7}.\label{tab.9}}
\begin{tabular}{lll}
\hline\hline
     Meson  &     $\Delta_{HF}$  & $\Delta_{HF}(exp)$\\
\hline
 $D(2S)$  &        $72\pm 3$   &    abs\\

$D_s(2S)$  &     $72\pm 3$   & abs \\

 $c\bar c(1S)$ & 93.7 &    $ 116.6\pm 1.2$\\

 $c\bar c(2S)$&       47.9  &   $49\pm 4$\\
\hline\hline
\end{tabular}
\end{table}

As seen from Table IX,  the HFS of the $D(2S)$, $D_s(2S)$ mesons,
$\sim 70$ MeV, are two times smaller than those for the ground
states, i.e. for these states a picture is similar to that in
charmonium, when the HFS for the $2S$ states is 2.3 times smaller
than the $J/\psi-\eta_c(1S)$ mass difference.

For the $D(2S)$, $D_s(2S)$ mesons our HFS appear to be
significantly smaller than the $D_s(2\,{}^3S_1)-D_s(2\,{}^1S_0)$
mass difference, predicted in \cite{ref.45}, where it is equal
151~MeV, and even larger $\Delta_{HF}(D(2S))=188$ MeV and
$\Delta_{HF}(D_s(2S))=192$ MeV were obtained in \cite{ref.46}. On
the contrary, in our approach the HFS for the $2S$ states have
appeared to be rather close to those from the GI paper
\cite{ref.1}, although in a static potential our and their sets of
parameters are very much different, with an exception of the value
of the string tension, equal $0.18$~GeV$^2$ in both calculations.
Comparison our results and predictions from Refs. \cite{ref.1},
\cite{ref.45}, \cite{ref.46}, where different relativistic models
used, are presented in Table X. \\

\begin{table}
\caption{The masses of the singlet and triplet $2S$ states
(in~GeV) for $D(2S)$ and $D_s(2S)$
($\alpha_{HF}=0.31$).\label{tab.10}}
\begin{tabular}{lllll}
\hline\hline
State & this paper &  GI \cite{ref.1} & RRS \cite{ref.45}& MMS \cite{ref.46}\\
\hline
$D(2\,{}^1S_0)$&  2,570 & 2.58&  abs  & 2.483\\
$D(2\,{}^3S_1)$& 2.642&    2.64& abs&  2.671\\
$D_s(2\,{}^1S_0)$& 2.664 &  2.67 & 2.486  &  2.563\\
$D_s(2\,{}^3S_1)$& 2.736 &2.73& 2.637 & 2.755 \\ \hline\hline
\end{tabular}
\end{table}

The masses, presented in Table X, were calculated with the use of
different relativistic models, in particular the Spinless Salpeter
Equation was exploited in \cite{ref.1}, \cite{ref.45}, and also in
our calculations here, with the static potential described in
Section 3. From Table X one can see that the singlet and triplet
masses for the $D(2S)$ and $D_s(2S)$ mesons appear to be very
close to each other in our calculations and in \cite{ref.1}. On
the contrary, in \cite{ref.45} and  \cite{ref.46} predicted HFS
for the $2S$ states are 2.0 and 2.6 times larger than in our
calculations, and due to this result, their masses of a singlet
state, $M(D_s(2\,{}^1S_0))$, is by $\sim 100$~MeV and $\sim
170$~MeV, respectively, lower than in \cite{ref.1} and in our
analyasis.

For the triplet $2S$ states differences in predicted masses are
not large and in \cite{ref.1}, \cite{ref.46}, and our calculations
$M(D^*(2S))$ and $M(D_s^*(2S))$ lie in the range 2.64-2.67~GeV and
2.73-2.75 GeV, respectively. These predictions are in good
agreement with the experimental mass, $M_{exp}(D_s^*(2S))=2710\pm
2\pm^{12}_7$ MeV \cite{ref.47}, \cite{ref.48}, while in
\cite{ref.45} predicted mass is by $\sim 70$ MeV smaller.

>From our analysis it follows that observation of the singlet
states, $D(2\,{}^1S_0)$ and $D_s(2\,{}^S_0)$, is crucially
important for understanding of spin-spin interaction in systems of
large sizes, in particular, it could clarify what is a
characteristic value of the strong coupling in HF interaction.

\section{Conclusions}

In our study we have used a conception of a universal HF
interaction and observed that

\begin{enumerate}
\item
In the $B_q$ mesons and bottomonium  a good agreement with
experimental HFS are reached if a universal coupling
$\alpha_{HF}=0.310$ is used in the HF potential.

\item
Just with the same coupling, $\alpha_{HF}=0.310$, the
$\psi(3686)-\eta_c(2S)$ mass splitting appears to be in agreement
with experiment.

\item
Calculated here mass splitting, $M(B_c^*)-M(B_c)=57.9(6)$~MeV,
gives the mass of unobserved yet $B_c^*$ meson, $M(B_c^*)=6.334\pm
5$~MeV. Our  HFS is close to that in full QCD calculations,
$\Delta_{HF}(B_c)=53(7)$~MeV \cite{ref.21}.

\item
In bottomonium  a full agreement with experimental mass of
$\eta_b(1S)$ is reached only if in the static potential $n_f=5$ is
used, giving $\Delta_{HF}(b\bar b)=71.1$~MeV.  For $n_f=4$ and
$n_f=3$ calculated HFS are smaller, being equal 63.4~MeV and
58.7~MeV, in agreement with the lattice results for $n_f=3$.

\item
With $\alpha_{HF}=0.310$ for the bottomonium $2S$ and $3S$
states the HFS, equal 36(1)~MeV and 27(1)~MeV, are predicted.

\item
The following masses of excited $B_q(2S)$ states are predicted:
$M(B(2S))=5967$ ~MeV, $M(B^*(2S))=6001$~MeV,
$M(B_s(2S))=6040$~MeV, $M(B_s^*(2S))=6075$~MeV,
$M(B_c(2S))=6832$~MeV, $M(B_c^*(2S))=6870$~MeV.

\item
We predict that the mass differences for $D^*(2S)-D(2S)$,
$D_s^*(2S)-D_s(2S)$ are $\sim 72(3)$~MeV, being smaller than in
several other analysis.

\item
We expect that nonperturbative spin-spin potential gives not small
contribution, $\sim 15-30\%$ to the mass splittings for
$J/\psi-\eta_c(1S)$, $D^*(1S)-D(1S)$, and $D_s^*(1S)-D_s(1S)$.

\end{enumerate}

For singlet states, $D(2\,{}^1S_0)$ and $D_s(2\,{}^1S_0)$, one
cannot exclude that they have not small hadronic shifts due to
coupling to open channels; if such hadronic shifts are small, then
observation of their masses, close to the values $\sim
2.57(1)$~GeV and $2.66(1)$~GeV, can be considered as a crucial
test of a universal character of spin-spin interaction.

\begin{acknowledgments} This work is supported by the Grant
NSh-4961.2008.2.
\end{acknowledgments}

\vspace{1cm}

\setcounter{equation}{0}
\renewcommand{\theequation}{A.\arabic{equation}}

\section{Appendix A. The pole mass of a heavy quark}

The RSH  (\ref{7}) contains the pole mass of a heavy quark
$m_2=m_Q$ and of a lighter quark $m_1$. For a light quark, $m_u$
or $m_d$, its mass is taken equal 5~MeV , while for a $s$ quark
$m_s=200$~MeV is used as in \cite{ref.26}. The pole masses of
heavy quarks ($c,b$) are defined as in pQCD, when the pole mass is
expressed via the QCD current mass $\bar m_Q(\bar m_Q)$, entering
the QCD Lagrangian, and higher order corrections of the strong
coupling $\alpha_s$ \cite{ref.7}:
\begin{equation}
 m_Q (pole)=\bar m_Q(\bar m_Q)\left[ 1 +\frac{4\alpha_s(\bar m_Q)}{3\pi} +
 r_2~\left(\frac{\alpha_s}{\pi}\right)^2\right],
\label{A.1}
\end{equation}
where  the factor $r_2$,
 \begin{equation}
 r_2(n_f)=13.4434 - 1.0414\sum^{N_L}_{k=1}\left(1-\frac{4\bar
 m_{Q_k}}{3\bar m_Q}\right),
 \label{A.2}
 \end{equation}
depends on a number of flavors $n_f$ through the sum, which goes
from $k=1$ up to $N_L=n_f-1$. Due to this term the pole mass
appears to be slightly different for different $n_f$. Here in our
calculations we take $m_b(n_f=5)=4.823(3)$~GeV and
$m_b(n_f=4)=4.79$~GeV; a small difference between them lies within
a theoretical error, present in the current mass, $m_b(\bar
m_b)=4.20\pm 0.07$~GeV \cite{ref.7}. The $m_b(pole)$ used here
correspond to the value of the current mass $\bar m_b(\bar
m_b)=4.210\pm 0.015$~GeV, which is within the conventional number
\cite{ref.7}.

For a $c$ quark the pole mass $m_c(n_f=4)=1.41$~GeV is used for
charmonium, and the $B_c$, $D_s$ mesons; it corresponds to the
conventional value, $\bar m_c=1.24\pm 0.09$~GeV \cite{ref.7}. For
the $D$ mesons $m_c(pole)=1.39$~GeV is used.

\section{Appendix B. The quark self-energy contribution to a meson mass}

The nonperturbative quark self-energy (SE) contribution
$\Delta_{SE}$ to a meson mass $M(nL)$ was calculated in FCM  \cite
{ref.32}, where the meson Green's function was defined in  a
gauge-invariant way. This correction,
\begin{equation}
\Delta_{SE}= - \frac{1.5 \sigma \eta(q)}{\pi \omega_q} \label{B.1}
\end{equation}
is negative, proportional to the string tension $\sigma$, and  a
quark kinetic energy as $ \omega_q^{-1}$, which is rather large
for a light quark and small for a heavy quark. In (\ref{B.1}) the
factor $\eta_q$ (a number) depends on a quark mass: $\eta_n = 1.0$
for a light quark; $\eta_s\simeq 0.80$ for a  $s$ quark;
$\eta_c\simeq 0.40$ for a $c$ quark, and $\eta_b\simeq 0.2$ for a
$b$ quark \cite{ref.48}, \cite{ref.32}. Notice that the SE term
(\ref{B.1}) contains correct coefficient 1.5, instead of the
coefficient 2.0 in \cite{ref.32}; the reasons for a change of this
number is discussed in \cite{ref.49}.

For a $b$-quark the SE contribution is small ($\leq (-1)$~MeV) and
can be neglected; for a $c$ quark its value is $\sim (-20)$~MeV
and it is convenient  to include this small correction via a
redefinition of a pole mass. For the ground states of heavy-light
mesons the SE contribution, which comes from a light, is rather
large, being $\sim (-140)$~MeV, and a bit smaller, $\sim
-90)$~MeV, for a $s$ quark. For higher states of heavy-light
mesons the SE contributions are smaller, because for them the
kinetic energy of a light ($s$ quark), present in (\ref{B.1}), is larger.\\

\end{document}